\documentclass{ieeeaccess}
\usepackage{cite}
\usepackage{amsmath,amssymb,amsfonts}
\usepackage{algorithmic}
\usepackage{graphicx}
\usepackage{textcomp}
\usepackage{url}
\usepackage{xcolor}
\usepackage{caption}
\DeclareCaptionFont{ieeeblue}{\color{accessblue}}
\DeclareCaptionLabelFormat{myformat}{\figcapfont{\textbf{#1}\textbf{~#2}}}
\def\BibTeX{{\rm B\kern-.05em{\sc i\kern-.025em b}\kern-.08em
    T\kern-.1667em\lower.7ex\hbox{E}\kern-.125emX}}
\begin{document}
\history{Date of publication xxxx 00, 0000, date of current version xxxx 00, 0000.}
\doi{10.1109/ACCESS.2019.DOI}

\title{Joint Activity Recognition and Indoor Localization with WiFi Fingerprints}
\author{
\uppercase{Fei Wang}\authorrefmark{1,}\authorrefmark{2},
\uppercase{Jianwei Feng\authorrefmark{2},}
\uppercase{Yinliang Zhao\authorrefmark{1*}, }
\uppercase{Xiaobin Zhang\authorrefmark{1},}
\uppercase{Shiyuan Zhang\authorrefmark{1}, and}
\uppercase{Jinsong Han\authorrefmark{3,1}.}
}
\address[1]{Department of Computer Science and Technology, Xi'an Jiaotong University, Xi'an, Shaanxi 710049 China}
\address[2]{The Robotics Institute, Carnegie Mellon University, Pittsburgh, PA 
15213 USA}
\address[3]{Institute of Cyberspace Research, Zhejiang University, Hangzhou, Zhejiang 
310058 China}


\markboth
{Fei Wang \headeretal: Joint Activity Recognition and Indoor Localization with WiFi Fingerprints}
{Fei Wang \headeretal: Joint Activity Recognition and Indoor Localization with WiFi Fingerprints}

\corresp{Corresponding author: Yinliang Zhao (e-mail: yinliangzhao1960@gmail.com).}

\begin{abstract}
Recent years have witnessed the rapid development in the research topic of WiFi sensing that automatically senses human with commercial WiFi devices. This work falls into two major categories, i.e., the activity recognition and the indoor localization. The former work utilizes WiFi devices to recognize human daily activities such as smoking, walking, and dancing. The latter one, indoor localization, can be used for indoor navigation, location-based services, and through-wall surveillance. The key rationale behind this type of work is that people behaviors can influence the WiFi signal propagation and introduce specific patterns into WiFi signals, called WiFi fingerprints, which can be further explored to identify human activities and locations. 
In this paper, we propose a novel deep learning framework for joint activity recognition and indoor localization task using WiFi Channel State Information~(CSI) fingerprints. More precisely, we develop a system running standard IEEE 802.11n WiFi protocol, and collect more than 1400 CSI fingerprints on 6 activities at 16 indoor locations. Then we propose a dual-task convolutional neural network with 1-dimensional convolutional layers for the joint task of activity recognition and indoor localization. Experimental results and ablation study show that our approach achieves good performances in this joint WiFi sensing task. Data and code have been  made publicly available at~{\color{blue}\url{https://github.com/geekfeiw/apl}}.

\end{abstract}


\titlepgskip=-15pt

\maketitle

\section{Introduction}\label{sec:introduction}

Channel State Information of WiFi devices have been extensively explored for human sensing tasks such as activity recognition~\cite{wang2014eyes, wang2015understanding, palipana2018falldefi,abdelnasser2015wigest}, gesture recognition~\cite{tian2018wicatch,li2016wifinger,ali2015keystroke,al2016wiger}, indoor localization~\cite{kotaru2015spotfi,li2016dynamic,kotaru2017position,xie2015precise}, and health-care applications~\cite{wang2017wifall,fang2016bodyscan,wang2017rt,wang2017tensorbeat,wang2016human,zheng2016smokey,liu2015contactless}. This prosperity benefits from several special properties of WiFi, including the ubiquitous deployment of commercial WiFi devices, the robustness to lighting condition and occlusion which overcomes limitation of cameras, and the non-intrusiveness sensing which requires no user's extra effort.

Though there is abundant work on the specific aforementioned WiFi human sensing task~\cite{qian2017inferring,li2016wifinger,ali2015keystroke,palipana2018falldefi,abdelnasser2015wigest,wang2017wifall,fang2016bodyscan,wang2017rt,wang2017tensorbeat,wang2016human,zheng2016smokey}, little work aims at completing the joint task of activity recognition and indoor localization. Carrying out the joint task would breed numerous useful human-computer interaction applications. For example, in a smart home with Internet-of-Things~(IoT) devices~\cite{atzori2010internet,gubbi2013internet}, the devices could precisely response differently to the same gesture command based on user's location. More specifically, the user can use the gesture of `hand down' to turn down the television in front of her, whereas she can also use the same gesture to lower the air conditioner's temperature when standing close to the air conditioner. To our best knowledge,  MultiTrack~\cite{tan2019multitrack} is the only work that enables indoor localization and activity recognition jointly, however it requires high-end hardware modification for ultra-wide band WiFi~(over 600MHz).

The joint task can be summarized as the following two folds. (1) Recognizing activities conducted at different locations. (2) Localizing the user by the activities. However, there are two major challenges lying in the way. The first challenge is that WiFi fingerprint differs even when performing a same activity but at different locations, thus we need to look for a same representation for activities conducted at all locations. The second one is that WiFi fingerprints vary when performing different activities in one location, thus we have to explore distinguished features for each location from the fingerprint variances.

To conclude the above challenges formally, WiFi fingerprint, $W$, contains two components at the same time, activity category~($A$) and user location~($L$). We denote the WiFi fingerprint as $W(A,L)$. Joint activity recognition and indoor localization task aims to learn a function $f$, which is capable to classify activity categories ($f:W(A,L)\rightarrow A$) and to localize the user ($f:W(A,L)\rightarrow L$), simultaneously. Thus we formalize the joint task as $f:W(A,L)\rightarrow (A, L)$.

To this end, in the paper we propose a novel 1-dimensional Convolutional Neural Network~(C1D) including two branches, one for activity recognition and the other for indoor localization. To date, conventional 2-dimensional Convolutional Neural Networks~(C2D), which have brilliant ability to learn features from raw data, boost the development of computer vision~\cite{he2016deep,krizhevsky2012imagenet, simonyan2014very,he2017mask,kurakin2016adversarial,CycleGAN2017}, robotics~\cite{li2016vehicle,wu2018squeezeseg,wang2018embedding,sun2018scene}, machinery~\cite{xia2018fault,jia2018deep,guo2018deep}, etc. Unlike C2D that processes 2D spatial data such as images, C1D is capable to process 1D temporal data. For temporal WiFi fingerprints, we design a C1D based on the ResNet~\cite{he2016deep} to carry out the joint task of activity recognition and indoor localization.

To evaluate our proposed approach, we implement the standard IEEE 802.11n protocol in two universal software radio peripheral~(USRP) sets, Ettus N210\footnote{\url{https://www.ettus.com/all-products/un210-kit/}}, where one Ettus N210 broadcasts WiFi signals and the other parses Channel State Information~(CSI) fingerprints of WiFi for joint task. We define 6 hand gestures for potential human-computer interaction applications, namely, hand up, hand down, hand left, hand right, hand circle and hand cross. One volunteer repeats these activities 15 times at each location~(16 locations in all) and forms a dataset with 1394 samples (after excluding the invalid data). We evaluate our proposed C1D on this dataset and present the results with several metrics such as  confusion matrix, F1 scores, convolutional feature maps, etc. Experimental results show our proposed C1D achieves a very promising performance in the joint task. We summarized our contributions as follows.

1. We first propose and achieve the joint task of activity recognition and indoor localization, which enable practical user gesture control in smart homes for human-computer interaction applications. 

2. We novelly view CSI fingerprints as time series with channel dimension and time dimension, apply a advanced 1-dimensional Convolutional Neural Network that sweeps along the time dimension of the CSI fingerprints, and achieve the joint task of activity recognition and indoor localization.

3. We implement IEEE 802.11n protocol in two USRP sets and build a dataset specifically for the joint task. We evaluate the performance of proposed deep network on this dataset and fully analyze the results.

\section{Related Work}

\subsection{CSI Fingerprints}
CSI fingerprints of WiFi have been widely utilized for activity recognition~\cite{li2016wifinger,ali2015keystroke,palipana2018falldefi,abdelnasser2015wigest,wang2017wifall,wang2017rt} and indoor localization~\cite{wang2017csi,zhang2018enhancement,ibrahim2018cnn,wang2017cifi}. As for activity recognition, in~\cite{palipana2018falldefi,wang2017wifall,wang2017rt}, CSI fingerprints are used to detect user falling especially for the elderly-care system. In~\cite{ali2015keystroke}, CSI fingerprints are used to infer user keystroke. Further in~\cite{li2016csi}, researchers find that CSI fingerprints can reveal people's typing when they use smart phones in public WiFi. In~\cite{li2016wifinger,wang2018csi} 
CSI fingerprints are designed for hand sign recognition 
for human-computer interactions. As for indoor localization, \cite{wang2017csi,zhang2018enhancement,ibrahim2018cnn,wang2017cifi} collect CSI fingerprints corresponding to people locations, and train classifiers to localize people with collected CSI fingerprints. To our best knowledge, there is no work on joint activity recognition and indoor localization, which is very useful in controlling different smart devices at different locations with a set of pre-defined activities. We achieve this task by a dual-branch Convolutional Neural Network.

\subsection{CSI Fingerprints Classification}
There exist three popular approaches in CSI fingerprints classification. (1) Hand-crafted features + Support Vector Machine~(SVM)~\cite{suykens1999least}: \cite{wang2017wifall,wang2017rt} apply statistical values of CSI time-series such as the mean, maximum, minimum, entropy, etc., as features to train SVM with kernel methods for CSI fingerprints classification. This approach requires expertise in designing features, which is even much harder on joint activity recognition and indoor localization. (2) Dynamic Time Wrapping~(DTW) + k Nearest Neighbors~(kNN): \cite{palipana2018falldefi,li2016wifinger,wang2018csi} first build a dataset with CSI fingerprints. When classifying a test CSI sample, this approach requires computing all distances between the test sample and all samples in the dataset, which is time-consuming compared to pre-training a classifier first. (3) Deep learning: \cite{wang2017csi,wang2017cifi} utilizes deep Boltzmann Machine~(DBM) to do indoor localization. However DBM relies heavily on careful design and tricks to converge. \cite{zhang2018enhancement,ibrahim2018cnn} apply 3-5 convolutional layers on activity recognition. In general, the shortage in the depth limits the performance. \cite{wang2018csi} utilizes ResNet~\cite{he2016deep} and Inception~\cite{simonyan2014very} to categorize CSI fingerprints, whereas it only handles single moment CSI, i.e., rather than handling temporal CSI fingerprints. In this paper, we propose a ResNet-based Convolutional Neural Network to do CSI fingerprints classification.

\subsection{1D Convolutional Neural Network}
Conventional Convolutional Neural Network~(C2D)~\cite{krizhevsky2012imagenet,simonyan2014very,he2016deep} are designed for 2D inputs such as images. C2D applies 2D convolutional kernels to sweep along the width and height of an image to capture its semantic and structural information for image classification~\cite{krizhevsky2012imagenet,simonyan2014very,he2016deep}, object detection~\cite{ren2015faster}, instance segmentation~\cite{he2017mask}, etc. In \cite{ji20133d,hara2018can}, researchers apply 3D CNN~(C3D) on video data, which sweeps along the width, height, and time of the video to capture information both in spatial and in temporal. In this paper, we apply 1D convolutional kernels to sweep along the time axis of the CSI fingerprint series to capture the temporal information of CSI fingerprints, which works well in the joint task of activity recognition and indoor localization.

\Figure[t!](topskip=0pt, botskip=0pt, midskip=0pt)[width=0.97\textwidth]{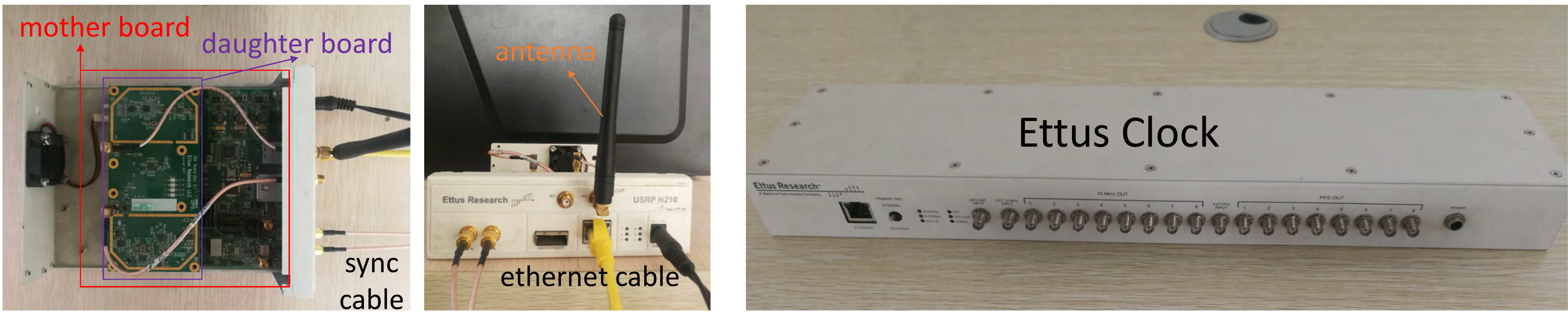}
{Main hardwares: Ettus USRP N210 and Ettus Clock.\label{fig:hardware}}

\Figure[t!](topskip=0pt, botskip=0pt, midskip=0pt)[width=0.48\textwidth]{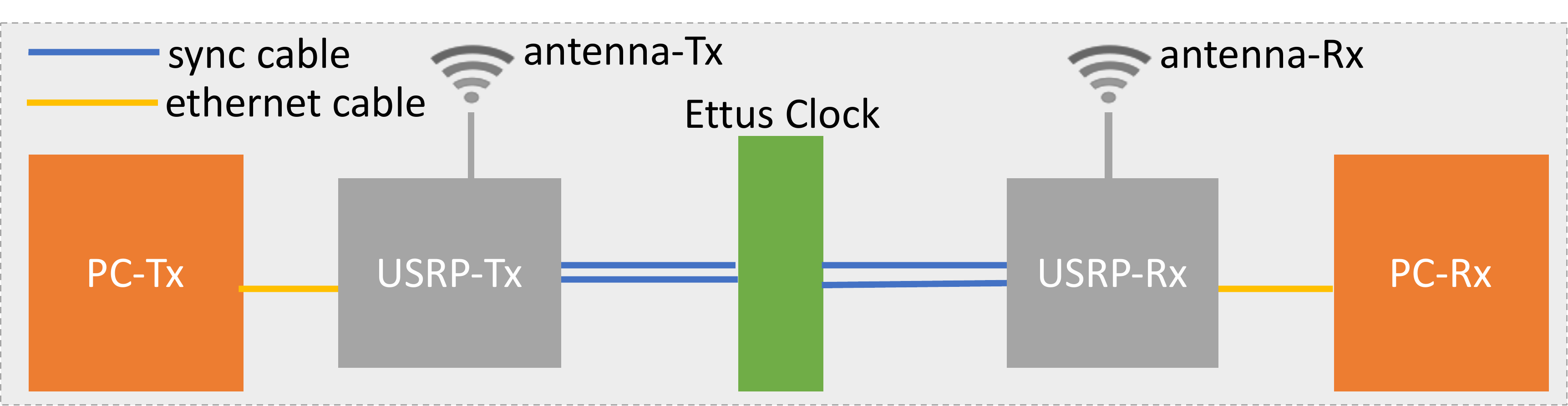}
{System framework. The system contains two sets of personal computers and USRPs, which work as the WiFi transmitter and receiver, respectively. An Ettus clock synchronizes the two sets.\label{fig:block}}
\section{Data Collection}

\subsection{Hardware}

We implement the standard IEEE 802.11n protocol in two universal software radio peripherals~(USRPs) to collect CSI fingerprints. As shown in FIGURE~\ref{fig:hardware}, the first two figures are the top view and front view of the USRP~(Etuss N201), respectively. The USRP is mainly composed of a mother board, a daughter board and a WiFi antenna, which is used to broadcast or receive WiFi signals under the control of GNU Radio\footnote{\url{https://www.gnuradio.org/}}. The details are listed below. Meanwhile, the assembling diagram is shown in FIGURE~\ref{fig:block}. 

1. Etuss N210s: A hardware with field programmable gate array~(FPGA) that can be embedded IEEE 802.11n protocol to send and receive WiFi packages for CSI fingerprints. 

2. Etuss Clock\footnote{\url{https://www.ettus.com/all-products/OctoClock-G/}} and synchronization cables:  Synchronizing N210s with GPS clock to avoid a WiFi phase shifting caused by the clock differences between two N210s.  

3. Antennas: To broadcast or receive WiFi signals under the control of GNU Radio\footnote{\url{https://www.gnuradio.org/}}

4. Computers and Ethernet cables: To control N210s when are set in a same local area network as N210s.

\Figure[t!](topskip=0pt, botskip=0pt, midskip=0pt)[width=0.48\textwidth]{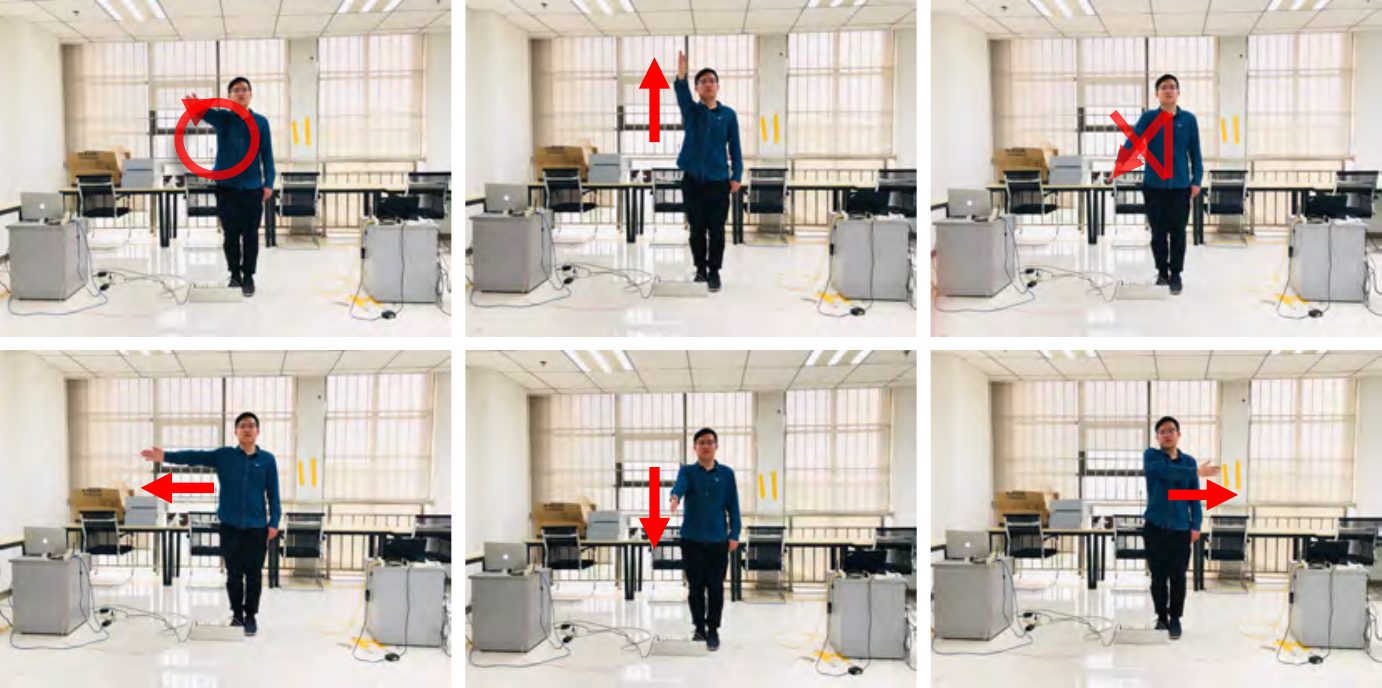}
{Six gesture commands mainly for human-computer interaction applications in smart home, i.e., hand circle, hand up, hand cross, hand left, hand down, and hand right. \label{fig:doing}}

\Figure[t!](topskip=0pt, botskip=0pt, midskip=0pt)[width=0.48\textwidth]{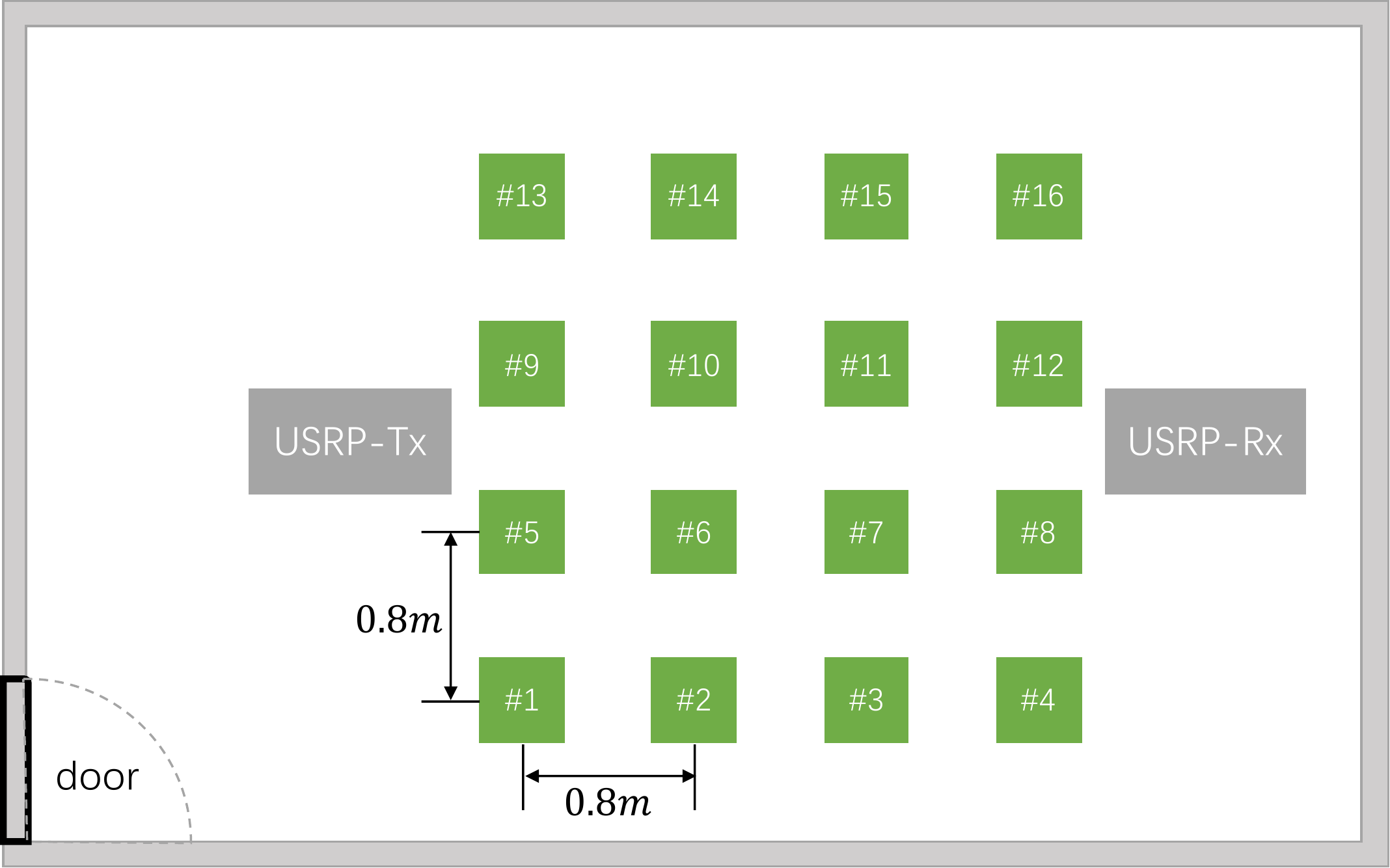}
{One volunteer does activities at 16 locations. \label{fig:location}}

\Figure[t!](topskip=0pt, botskip=0pt, midskip=0pt)[width=1.0\textwidth]{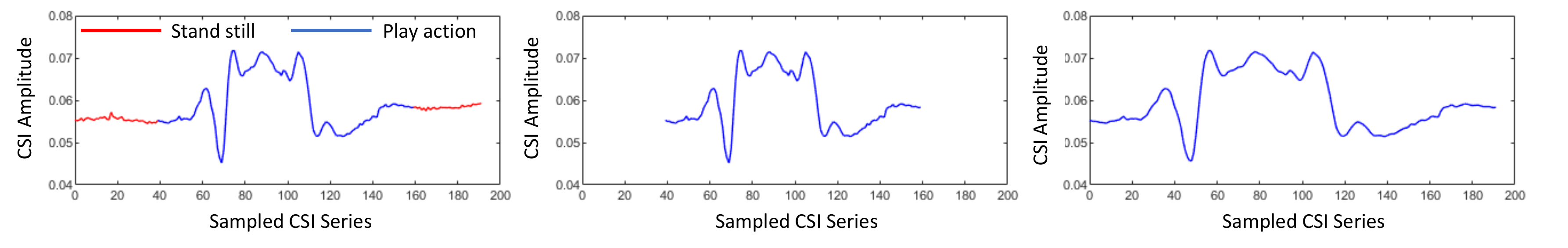}
{Signal preprocessing. we manually annotate the action duration~(leftmost), split  CSI samples during action~(middle), and upsample the splitted CSI series to be size of 192~(rightmost). (take the 29th subcarrier series when the volunteer plays a `circle' action at \#2 position for example.)\label{fig:upsample}}

\Figure[t!](topskip=0pt, botskip=0pt, midskip=0pt)[width=0.48\textwidth]{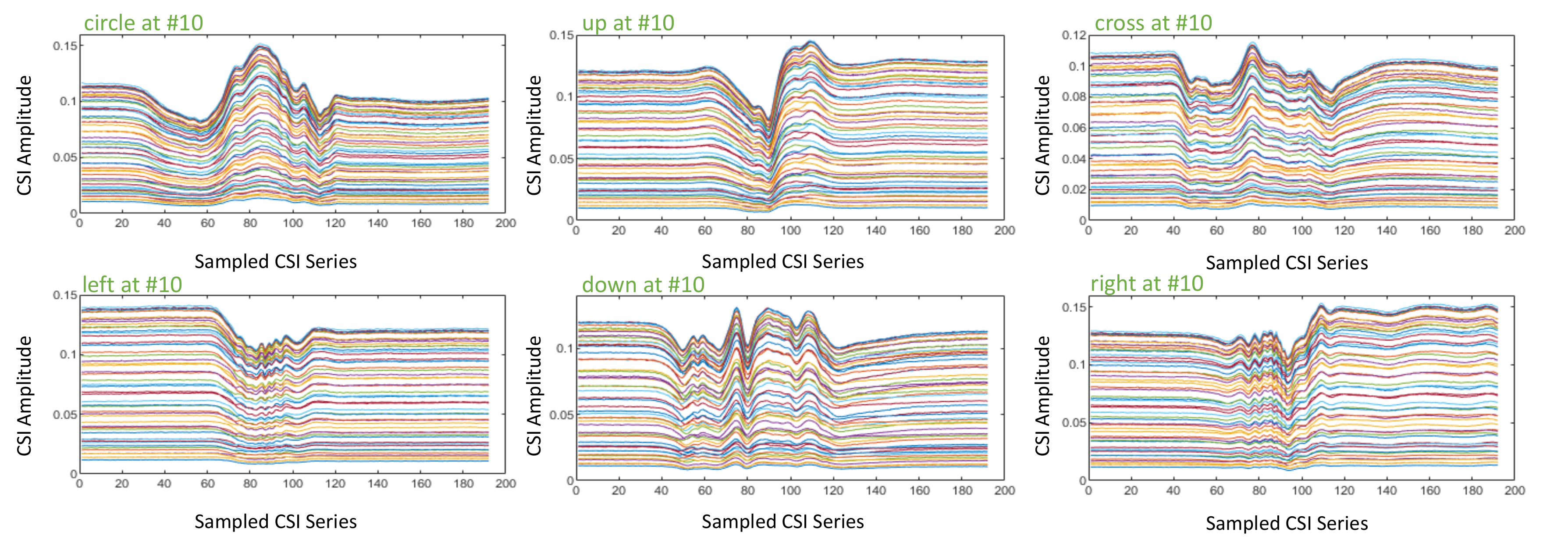}
{CSI fingerprint samples of 6 activities on \#10 position. \label{fig:activity}}

\Figure[t!](topskip=0pt, botskip=0pt, midskip=0pt)[width=0.48\textwidth]{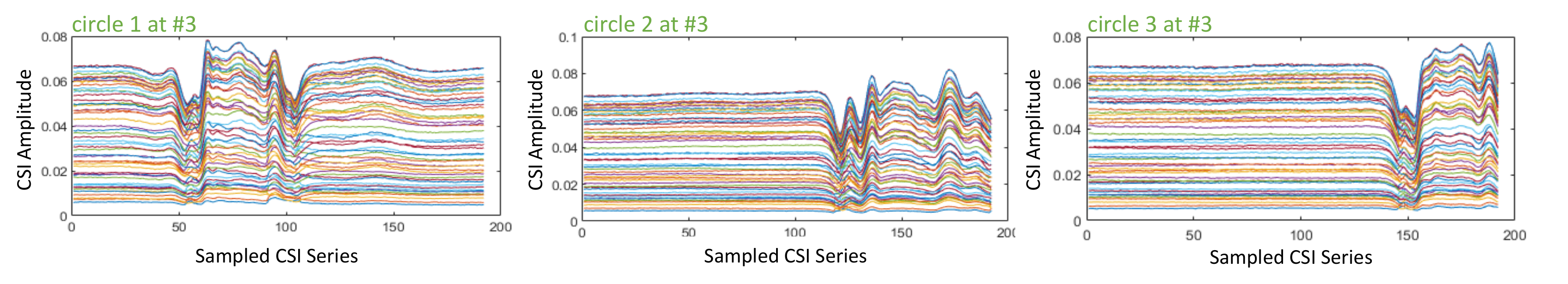}
{Do `circle' at the \#3 position. CSI samples vary in profiles. CSI may be partly captured because of late action start.  \label{fig:circle46}}

\Figure[t!](topskip=0pt, botskip=0pt, midskip=0pt)[width=0.48\textwidth]{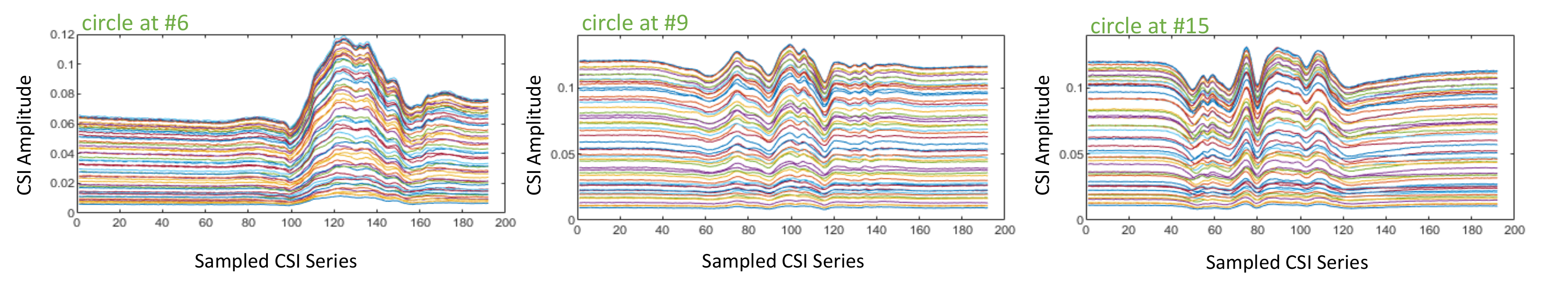}
{Do `circle' at the \#6~(left), the \#9~(middle), and the \#15 position. \label{fig:locations}}

\subsection{Activity and Location}\label{sec:actandloc}

We design 6 activities, namely, hand up, hand down, hand left, hand right, hand circle, and hand cross, for human-computer interaction applications, as shown in FIGURE~\ref{fig:doing}. This cluster of activities covers the majority of daily commands for smart Internet-of-Things home, where using cameras are not practical due to security and privacy concerns. Here we illustrate how our proposed activities work by the case of television. ``Hand up'' and ``hand down'' can be used to turn up and down the voice volume, respectively; ``Hand left'' and ``hand right'' indicate switching channels; ``Hand circle'' and ``hand cross'' are for CONFIRM command and CANCEL command, respectively. 

Besides recognizing activity in smart home, localizing the user when s/he is doing an activity is also crucial for the joint task. By combining user's activity and location together, we are able to infer user's intention more precisely and make it possible for users to control a range of smart devices with the same activity. For example, the user may want to communicate with the television when sitting on the sofa, whereas s/he probably needs to control the air conditioner~(AC) when standing in front of an AC. To make a proof-of-concept experiment, we collect CSI fingerprints when a volunteer does 6 activities (shown in FIGURE.~\ref{fig:doing}) at 16 locations in one room. We illustrate the spatial relationship in FIGURE.~\ref{fig:location}, where the 16 locations are selected evenly with a purpose to cover most central area of the room. The USRPs are fixedly placed besides the selected locations. In all, the volunteer repeats each of the 6 activities 15 times at 16 locations and forms a dataset with 1440 samples.

\subsection{CSI fingerprint analysis}
We visualize some samples varying in activities and locations to present the challenges of joint task of  activity recognition and indoor localization. FIGURE~\ref{fig:activity} shows CSI fingerprints when the volunteer does 6 activities at \#10 location, in which the x-axis is the sampling index~(time) and the y-axis is the amplitude of the CSI fingerprints. There are 52 time series in each CSI fingerprints, differed with 52 colors in FIGURE~\ref{fig:activity}. 52 is the number of orthogonal frequency division multiplexing~(OFDM)~\cite{nee2000ofdm} sub-carriers that carry data in parallel in WiFi protocol. FIGURE~\ref{fig:activity} demonstrates that CSI fingerprints vary when the user conducts 6 activities at a same location.

 FIGURE~\ref{fig:circle46} illustrates 3 CSI fingerprint samples when the volunteer carries out ``hand circle'' at \#3 location. FIGURE~\ref{fig:circle46} shows that though the volunteer plays the same activity at the same location, CSI fingerprints are still very different in time-serial profile~(left and middle), and in the start point of the activity~(left and right). Besides, performing the same activity at different locations also largely varies CSI fingerprints as illustrated in FIGURE~\ref{fig:locations}, making it challenging to find out shared features for one activity at all locations.

\section{Methodology}

\subsection{Preprocessing}\label{sec:preprocess}

As shown in FIGURE~\ref{fig:block}, we have one transmitting antenna and one receiving antenna. As shown in FIGURE~\ref{fig:circle46}, the number of sub-carriers and the streams for each CSI fingerprint are 52 and 192, respectively. This setting makes 
each CSI fingerprint a matrix with size of $52\times 192$. Note that we only use CSI amplitude and ignore the CSI phase for the fusion of both is out of range of this paper. Because CSI fingerprints vary according to different activity start time and finish time, we manually annotate the activity duration to prepare useful signals for further use. We take the time series of 29th sub-carrier as the visualization example to show a duration annotation in FIGURE~\ref{fig:upsample}~(left). This annotating process enables us to directly use the segmented CSI fingerprints for the joint task. We then upsample the segmented CSI fingerprints to make them the same size using the linear interpolation~(in our experiment, the sizes of original and upsampled CSI fingerprints are both 192). One interpolated sample is shown in FIGURE~\ref{fig:upsample}~(right).

\subsection{1D Convolutional Neural Network}\label{sec:1d}
As illustrated in FIGURE~\ref{fig:activity}, FIGURE~\ref{fig:circle46} and FIGURE~\ref{fig:locations}, CSI fingerprints are time series with 52 sub-carriers. We denote it as $C\in R^{52\times t}$, where $t$ is for sampled time, $R$ means \textit{Real} number. Thus it is essentially to do time-serial matching that to use CSI fingerprints for the joint task of activity recognition and indoor localization. Recall that Dynamic Time Warping~(DTW)~\cite{berndt1994using,bagnall2017great}, one of the most prevailing approaches for time-serial matching, takes both the values and the shapes of the time series into consideration to compute the distance between two series. However, to categorize one test series, DTW has to compute the distances of the series to all samples in training database, which is time-consuming. Some work utilizes hand-crafted statistics characteristics of the time series, e.g., the mean, variance, and entropy, to represents the values and shapes of the time series~\cite{wang2018continuous,wang2017wifall,wang2017rt}, which requires expertise to design and extract these characteristics, i.e., not generic. Thus to achieve the joint task, we aim to propose an approach that (1) can capture the values and the shapes of the CSI fingerprints, and (2) should be time efficient. Since Convolutional Neural Network approaches boost current pattern recognition applications due to the ability of learning powerful features directly from raw data. In this paper, we apply 1-dimensional Convolutional Neural Network~(C1D) on CSI fingerprints for joint task of activity recognition and indoor localization.

FIGURE~\ref{fig:1d}~(left) shows a 2-dimensional convolutional operation~(C2D) for spatial data such as images, and FIGURE~\ref{fig:1d}~(right) illustrates C1D for temporal inputs such as WiFi fingerprints. For a C2D, the input size is $7\times 7$, a 2D convolutional kernel sized $3\times 3$ sweeps along the width and height of the input with the stride of 2, and it leads to a result of $3\times 3$. With this sweeping operation, spatial information of the input, such as the object location in an image, can be captured. Differing from C2D, C1D only sweeps along the time axis and captures temporal information in $C$ (including values and shapes), which is highly correlated with the user activities. Besides, consulting the widely used method that C2D takes 3 (3 for RGB color channels) as the channel for image data, in this paper, we take the 52 of $C$ as the channel of C1D operation~(52 for 52 OFDM channels) for CSI fingerprints. Different to previous work that views the 2D CSI fingerprints as images with height and width, then applies C2D on the fingerprints~\cite{wang2018csi,zhou2018signal,zhang2018enhancement}, to our best knowledge, our approach is the first manner that views CSI fingerprints as time series with channel dimension and time dimension. More importantly, it makes advanced C1D can be directly applied on CSI fingerprints without bells and whistles.

\Figure[t!](topskip=0pt, botskip=0pt, midskip=0pt)[width=0.48\textwidth]{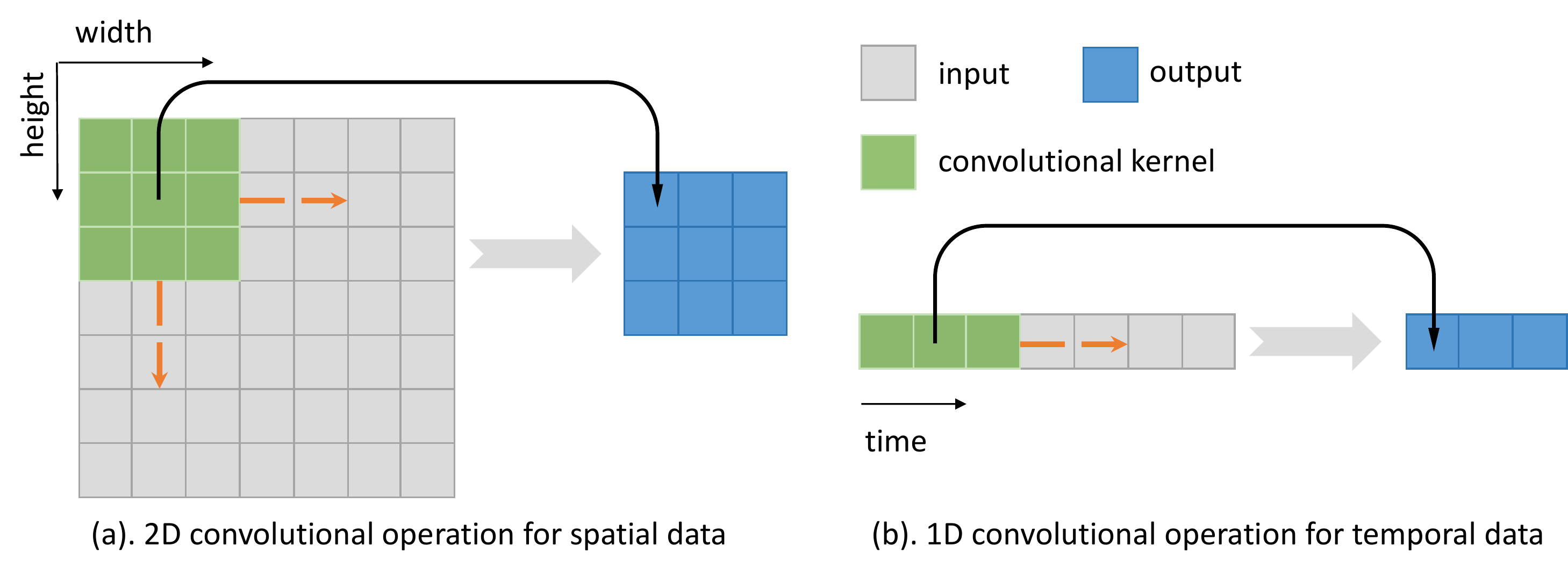}
{The difference between the 2D convolutional operation and the 1D convolutional operation.\label{fig:1d}}

\begin{figure}[t]
    \centering
    \includegraphics[width=1\linewidth]{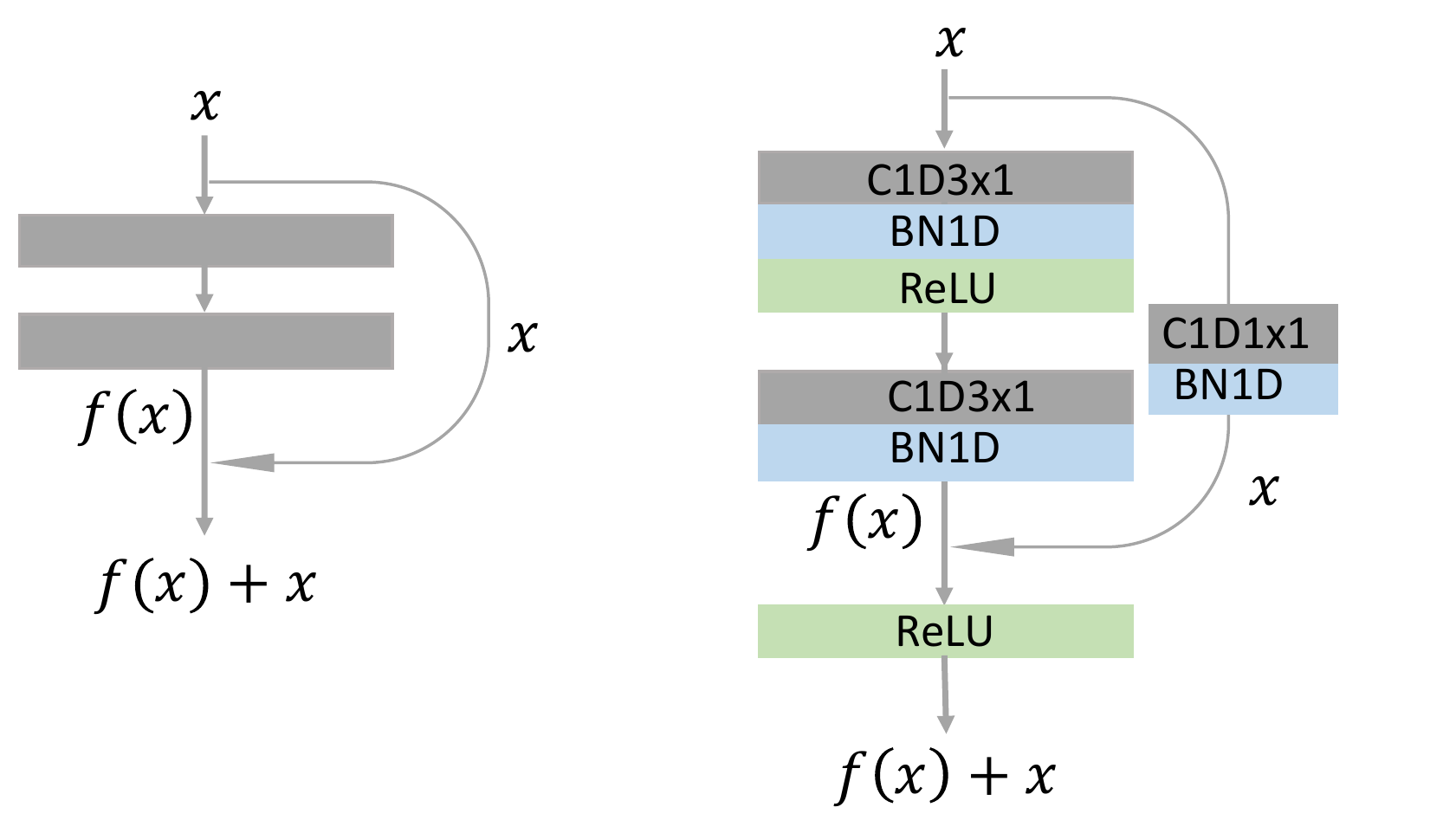}
    \caption{The basic residual block~(left) and a detailed implementation. `C1D3$\times$1' and `C1D1$\times$1' stand for the 1D convolutional operation with the size of $3\times1$ and $1\times 1$, respectively. `BN1D' is for 1D Batch Normalization. ReLU is for the Rectified Linear Unit activation function.}
    \label{fig:residual}
\end{figure}

\begin{figure}[t]
    \centering
    \includegraphics[width=0.35\textwidth]{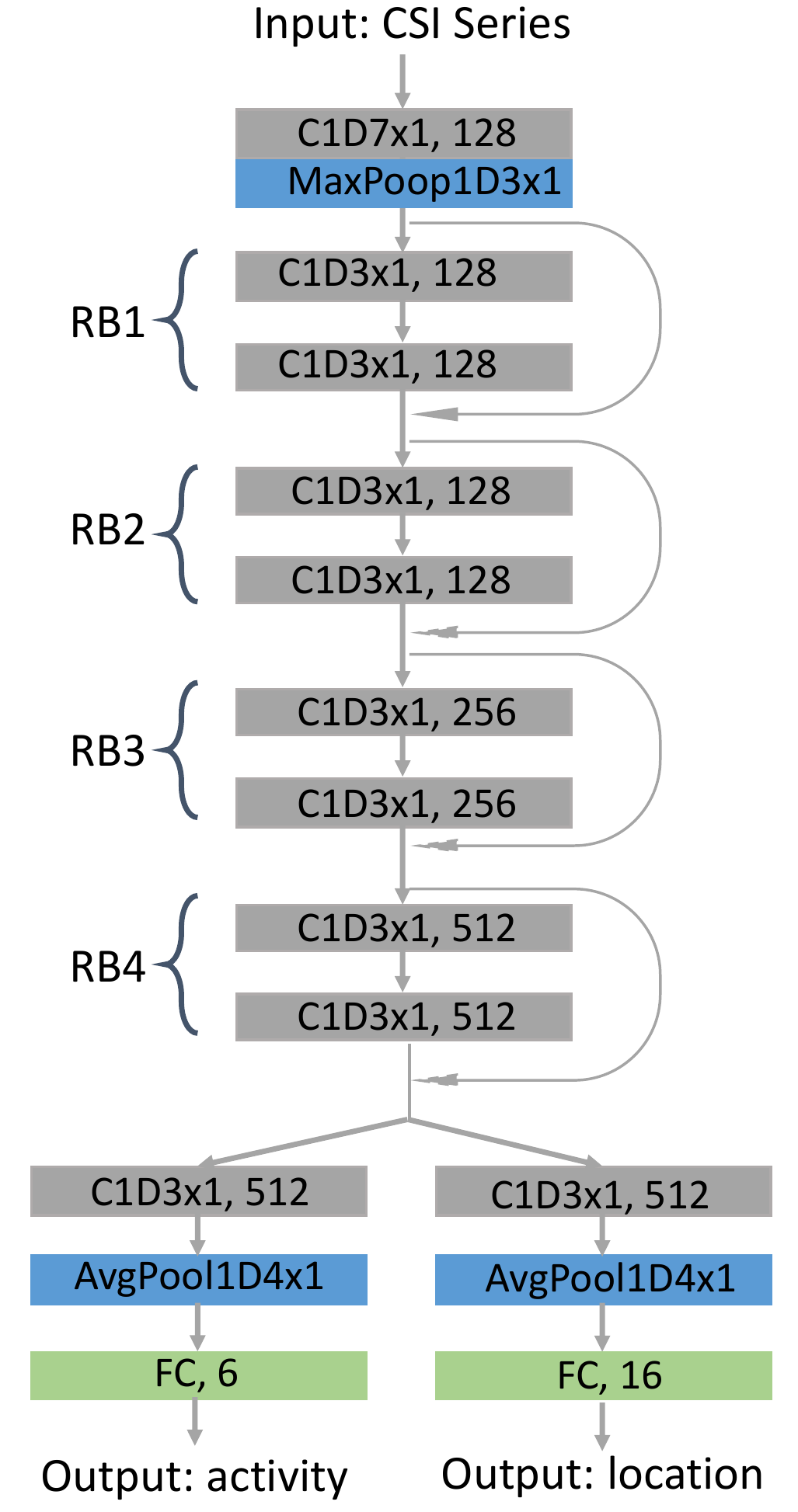}
    \caption{Deep framework. `RB' is the abbreviation of residual block. For all RBs having one residual shortcut, we term this framework as ResNet1D-[1,1,1,1]. }
    \label{fig:network}
\end{figure}

\subsection{Network Framework}
In computer vision community, ResNets~\cite{he2016deep} have been proved to be effective and advanced in many task, such as image classification~\cite{he2016deep}, object detection~\cite{ren2015faster}, instance segmentation~\cite{he2017mask}, etc. 
However the standard ResNets are implemented to process 2D inputs such as images with height and width, thus we re-implement a ResNet specifically for our temporal CSI fingerprints, termed as ResNet1D.

The main component of ResNet1D is the basic residual block as shown in FIGURE~\ref{fig:residual}~(left). We denote the input fingerprints as $x$, and the output as $y$. With two convolutional layers, $x$ becomes $f(x)$. With a shortcut link, $x$ becomes a part of $y$. Thus the two branches make the output of the basic residual block 
\begin{equation}
    y = f(x) + x.
\end{equation}
In FIGURE~\ref{fig:residual}~(right), we illustrate  our implementation in details. In the $f(x)$ branch, $x$ is scanned by two C1Ds with size of $3\times 1$~(`C1D$3\times$1'). Moreover, 1D batch normalization~\cite{ioffe2015batch}~(`BN1D') follows at each `C1D$3\times$1', and a Rectified Linear Unit activation function follows the first `C1D$3\times$1'. In the shortcut branch, $x$ is processed by a C1D with the size of $1\times1$~(`C1D$1\times$1') and a 1D batch normalization. The outputs of two branches,  $f(x)$ and $x$, may have difference in the size, making it unavailable to do element-wise addition between $f(x)$ and $x$. Therefore, the `C1D$1\times$1' in the shortcut branch is designed to make the sizes of $f(x)$ and $x$ the same.

\begin{figure*}[t]
    \centering
    \includegraphics[width=1\linewidth]{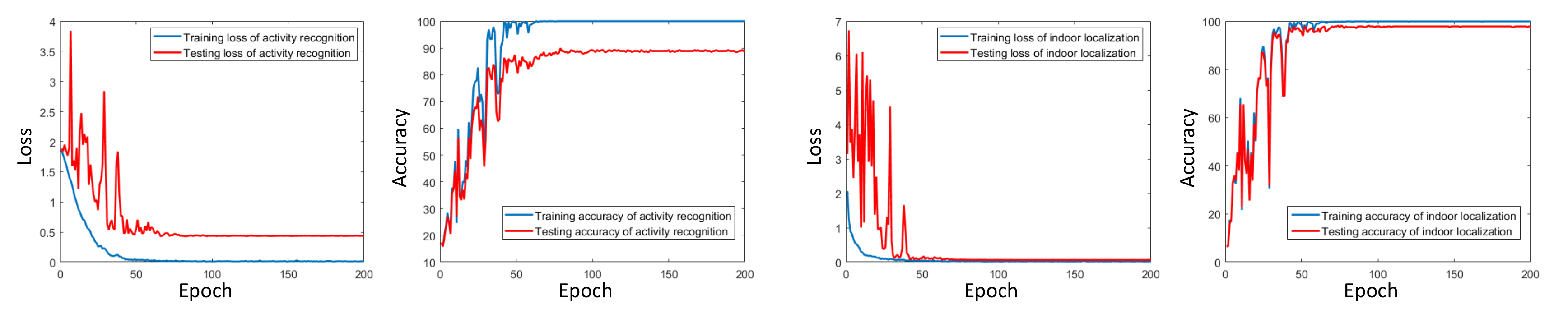}
    \caption{Learning curves and losses of activity recognition, and indoor localization.}
    \label{fig:curve}
\end{figure*}

Based on the basic residual block above, we build the ResNet1D as shown in FIGURE~\ref{fig:network}. The network takes CSI time series as input and predicts user activity and location in parallel as output.
The network contains 11 C1D layers, where 9 are shared, and each sub-task has one C1D independently, whose parameters are listed in FIGURE~\ref{fig:network}. Taking the first C1D, `C1D$7\times 1$, 128', as an example, the C1D is with the kernel size of $7\times 1$ and the output channel number of 128. Besides C1Ds, there is 1 max pooling operation following the first C1D and 2 average pooling operations following the last 2 C1Ds, respectively. Four residual blocks (RB1, RB2, RB3, and RB4) contribute the mainstream of the ResNet1D-[1,1,1,1]. Take the RB1 for example, it contains one basic residual block as shown in FIGURE.~\ref{fig:residual}, where the first convolutional layer is a 1D convolution with kernel size of $3\times 1$ and out-channel number of 128~(C1D$3\times1, 128$).
Note that we neglect the `BN1D and `ReLU' for simplification. Since this network contains 4 residual blocks~(RBs) and in each block there exists one basic residual block, we call it ResNet1D-[1,1,1,1]. Moreover, we use a fully-connected layer to predict each of the 6 activities with a separate score. The output of activity recognition is the activity with the highest score. Meanwhile we use a fully-connected layer to predict one location out of 16 locations where the activity is carried out.

It deserves mentioning that ResNet1D-[1,1,1,1] is an expansible framework by cascading multiple basic residual blocks in RBs. In the evaluation section, we will compare the performances of ResNet1D-[1,1,1,1], ResNet1D-[2,2,2,2], and ResNet1D-[3,3,6,3]~(the latter two RB settings are defaults in ResNet~\cite{he2016deep}).

\subsection{Loss function}

The loss, $L$, that optimizes ResNet1D is the sum of two sub-tasks, activity recognition and indoor localization. We term it as follows.
\begin{equation}
    L = L_{activity} + \lambda  L_{location}
    \label{equ:loss}
\end{equation} 
where $L_{activity} $ and $L_{location}$ are losses of activity recognition and indoor localization, respectively, $\lambda$ is to balance these two losses. 
Before computing $L_{activity} $ and $L_{location}$, we first normalize the prediction scores with SoftMax function, 
\begin{equation}
    s_{i}' = \frac{e^{s_i}}{ \sum_{j=1}^{K}e^{s_j} }, i \in [1,2,...,K],
    \label{equ:softmax}
\end{equation}
where $K$ is the categories of activities~($K=6$ for $L_{activity}$ and $K=16$ for $L_{location}$), $s_{i}$ and $s_{i}'$ are the predicted score and normalized score for the i-th activity, respectively. Using (\ref{equ:softmax}), all prediction scores are normalized to 0-1 range.

Then we apply the Cross Entropy Loss function on the normalized score to compute $L_{activity} $ as follows.
\begin{equation}
    L_{activity} = - log(s_{t}')
\end{equation}
where $s_{t}'$ means the normalized prediction score that belongs to (resulted from) t-th activity. With the same approach, indoor localization loss, $L_{location}$, can be computed. In our experiment, we assume activity recognition and indoor localization are of the same importance, thus we set $\lambda$ in (\ref{equ:loss}) as 1 to optimize ResNet1D-[1,1,1,1].

\subsection{Implementation}\label{sec:implementation}

We implement ResNet1D with Pytorch 1.0.0~\footnote{\url{https://pytorch.org/}} in a desktop that is with the Window 7 OS and one Nvidia Titan Xp GPU. The network is trained for 200 epochs by Adam optimizer~\cite{kingma2014adam}~with default settings~($\beta_1 =0.9$, $\beta_2=0.999$).
The mini-batch size is 128 and the initial learning rate is 0.005. The learning rate decays by 0.5 every 10 epochs. Before each epoch, all training data are shuffled.

\section{Evaluation}

\subsection{Dataset}

As \ref{sec:actandloc} described, our dataset involves 6 hand activities, i.e., hand up, hand down, hand left, hand right, hand circle and hand cross, that one user conducts at 16 locations. At each location, each activity is repeated for 15 times. Thus we collect totally $16\times 6 \times 15=1440$ samples. In~\ref{sec:preprocess}, we manually discard samples with extremely late start point to ensure data quality, leading to a final dataset with 1394 samples. We select one out of every five samples to build the test set~(278), and leave the remaining 1116 samples for the training set. Due to the test set evenly selected from all samples, the test condition is the same as the training condition. We then train the proposed deep networks with the training set and validate the networks on the test set.

\subsection{Learning curves}

\begin{table*}[t]
\centering
\begin{tabular}{l|l|l|l|l|l|l}
\hline
Activity & up   & down & left & right & circle & cross \\ \hline \hline
Precision            & 0.90  & 0.91 & 0.85 & 0.92  & 0.97   & 0.84  \\ \hline
Recall               & 0.98 & 0.89 & 0.83 & 0.92  & 0.77   & 0.89  \\ \hline
F1 score                   & 0.94 & 0.90  & 0.84 & 0.92  & 0.82   & 0.86  \\ \hline
\end{tabular}
\caption{Precision, recall, and F1 score of the activity recognition task.}
\label{tab:activity}
\end{table*}

\begin{table*}[t]
\centering
\begin{tabular}{l|l|l|l|l|l|l|l|l|l|l|l|l|l|l|l|l}
\hline
Location        & \#1 & \#2     & \#3 & \#4           & \#5 & \#6 & \#7           & \#8           & \#9           & \#10 & \#11 & \#12          & \#13          & \#14          & \#15 & \#16          \\ \hline\hline 
Precision & 1 & 0.88 & 1 & 0.87 & 1 & 1 & 1           & 1           & 1           & 1  & 1  & 0.89 & 0.95 & 0.89 & 1  & 0.85        \\ \hline
Recall    & 1 & 0.88 & 1 & 0.76 & 1 & 1 & 0.94 & 0.94 & 0.94      & 1  & 1  & 0.94 & 1           & 0.94 & 1  & 0.94 \\ \hline
F1 score       & 1 & 0.88 & 1 & 0.81      & 1 & 1 & 0.97 & 0.97 & 0.97 & 1  & 1  & 0.92 & 0.97 & 0.92 & 1  & 0.89 \\ \hline
\end{tabular}
\caption{Precision, recall, and F1 score of the indoor localization task.}
\label{tab:location}
\end{table*}

\begin{figure}[t]
    \centering
    \includegraphics[width=0.8\linewidth]{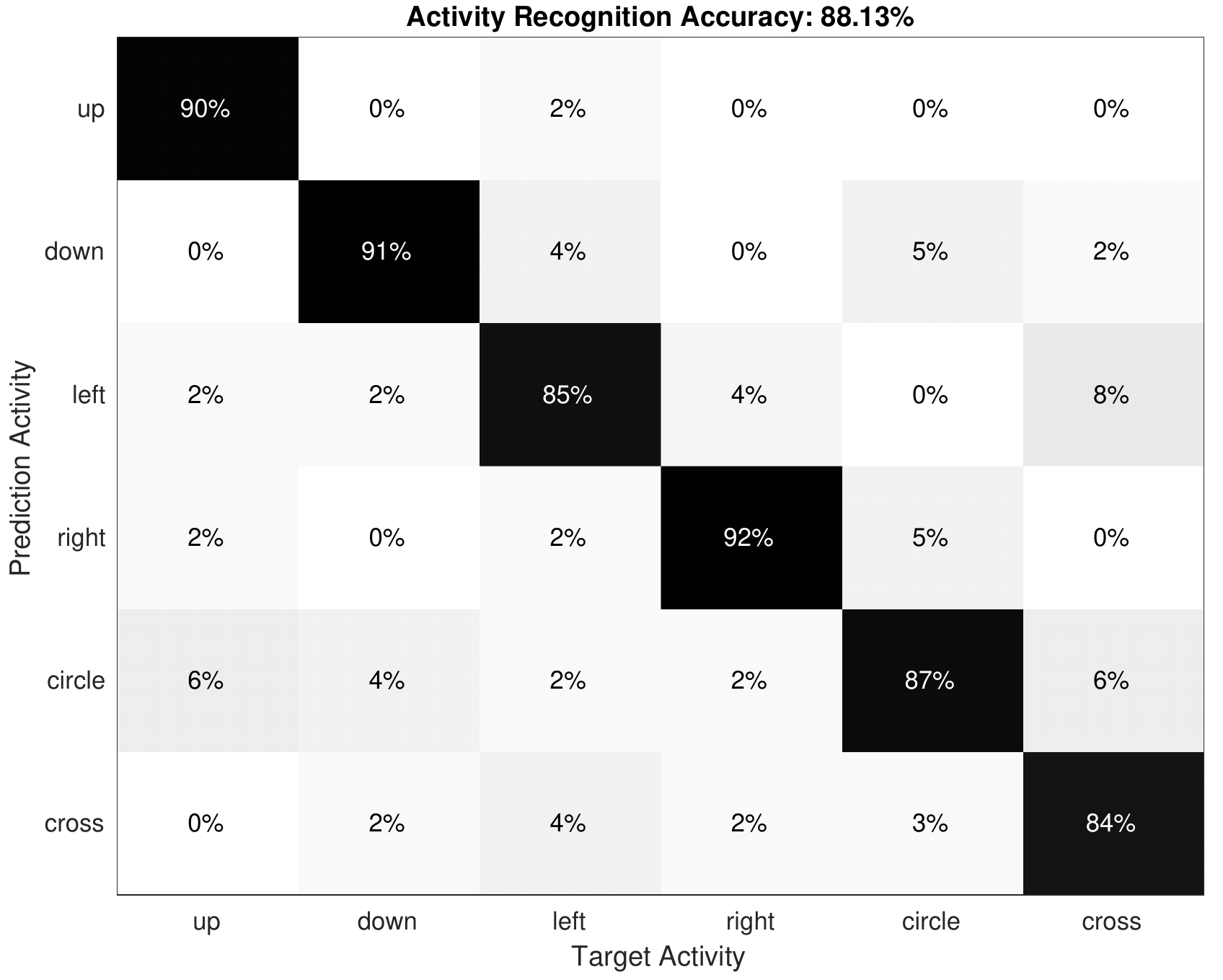}
    \caption{Confusion matrix of activity recognition.}
    \label{fig:conf_act}
\end{figure}

We display learning curves of loss and accuracy for the activity recognition and indoor localization in FIGURE~\ref{fig:curve}. In the loss curve of activity recognition~(1st subfigure), the training loss~(blue line) decreases gradually, and reaches a relatively low state around the 50th epoch. Whereas the test loss curve~(red line) wildly swings within the first 45 epochs, and gradually reaches to a steady state around the 75th epoch.

A phenomenon needs to be addressed is that though the training loss curve keeps relatively steady after the 50th epoch, the test loss still decreases when more training epochs are involved. We ascribe it to the process of shuffling training dataset before each training epoch, described in~\ref{sec:implementation}. The shuffle process makes the network, i.e., ResNet1D-[1,1,1,1], be optimized with different mini-batch samples in each epoch. After 
training loss curve reaching a steady condition, the shuffling process continuously generates (keeps generating) more mini-batch combinations and these combinations are continuously updating the network.

The accuracy curves of activity recognition are plotted in the 2nd sub-figure of FIGURE~\ref{fig:curve}, where the training curve~(blue line) reaches a steady condition around the 70th epoch, and the test curve reaches a steady condition around the 100th epoch. Similarly, we ascribe it to shuffling process as explained above. Besides, the learning curves of indoor localization are plotted in the 3rd and 4th sub-figures of FIGURE~\ref{fig:curve}. Comparing the learning curves between the activity recognition and indoor localization, we find that the task of indoor localization converges faster and achieves a better performance.

\begin{figure}[t]
    \centering
    \includegraphics[width=0.8\linewidth]{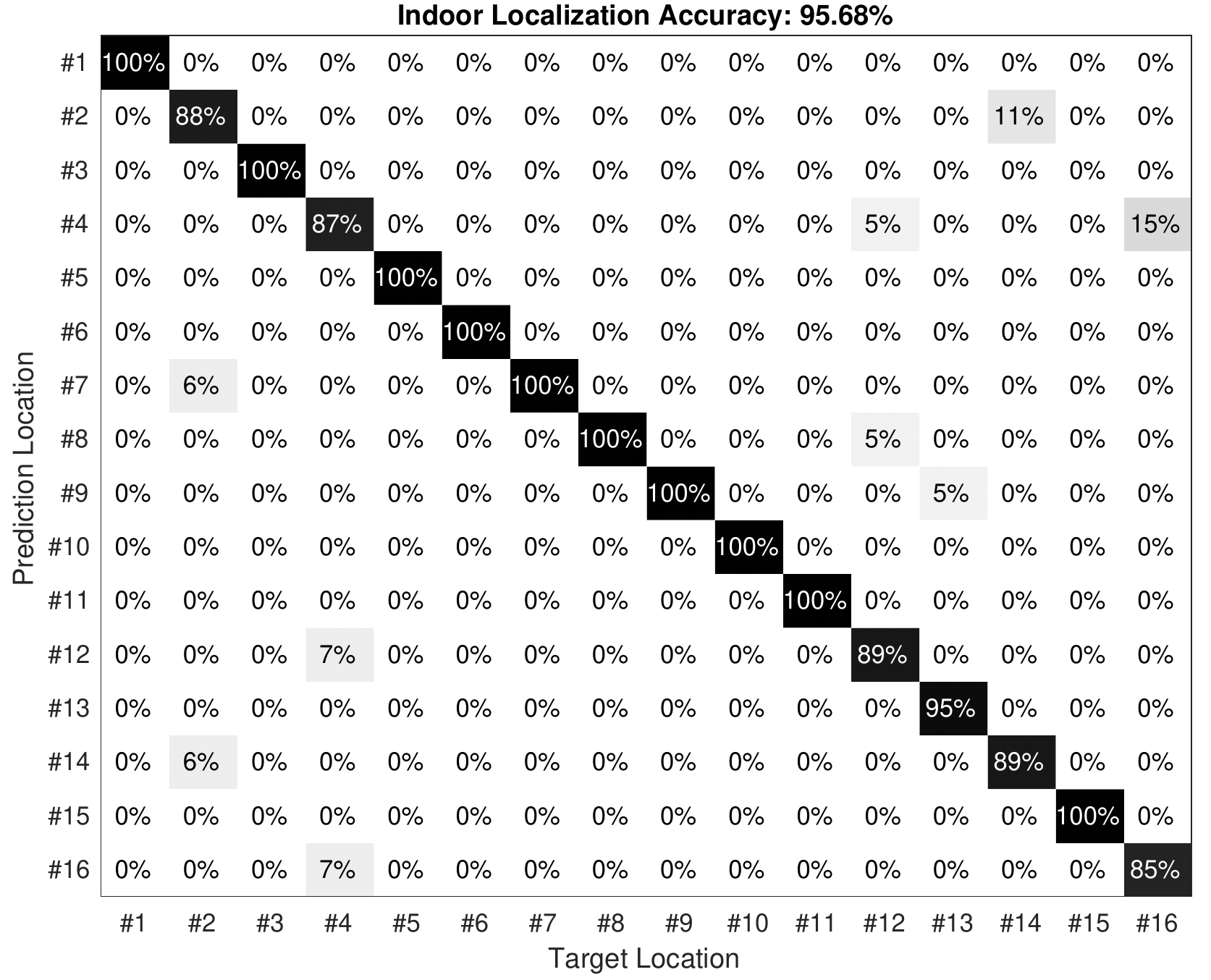}
    \caption{Confusion matrix of activity recognition.}
    \label{fig:conf_loc}
\end{figure}

\subsection{Quantitative results}

We demonstrate the quantitative results including confusion matrix, prediction accuracy, precision, recall, and F1 scores in the following section. 

The confusion matrix of ResNet1D-[1,1,1,1] on activity recognition and indoor localization are shown in FIGURE~\ref{fig:conf_act} and FIGURE~\ref{fig:conf_loc}, respectively. As shown in the two figures, we achieve a accuracy of 88.13\% for activity recognition and 95.68\% for indoor localization. The gap between two accuracies accords with learning curves. For activity recognition, the majority of mis-predictions happen at recognizing the gesture of hand cross. Precisely, ResNet1D-[1,1,1,1] wrongly predicts 8\% of hand cross to hand left, and wrongly predicts 6\% of hand cross to hand circle.
Meanwhile for indoor localization, a major error is wrongly predicting 15\% of \#16 location as \#4 location. Nevertheless, ResNet1D-[1,1,1,1] generally works well on both activity recognition and indoor localization.

We further compute the precision, recall, and F1 score from the confusion matrix, and list the results in Table~\ref{tab:activity} and Table~\ref{tab:location}. There exists a big gap between precision and recall for the activity of hand circle. A precision of 0.97 means ResNet1D-[1,1,1,1] effectively figures out (recognize) the hand circle activity, while a recall of 0.77 indicates that ResNet1D-[1,1,1,1] tends to categorize other activities into hand circle, decreasing the F1 score of hand circle to 0.82. Besides in Table~\ref{tab:location}, the lowest F1 score is on \#4 location prediction, 0.81, due to the low recall. In general, ResNet1D-[1,1,1,1] achieves very promising performances.

In addition, we computer average localization error~(ALE) in terms of meters by transferring localization labels via    
\begin{equation}\label{eq:meter}
ALE = \frac{1}{N}\sum_{i=1}^{N}\left \| Coor(P_i) -Coor(GT_i))  \right \|_2^2
\end{equation}
where $i$ and $N$ are the index and the number of test sample, respectively, ($N=278$); $Coor(P_i)$ is to transfer the predicted location label to the 2D spatial coordinates according to FIGURE.~\ref{fig:location}; $Coor(GT_i)$ is to transfer the ground-truth label to the 2D spatial coordinates; $\left \| \cdot  \right \|_2^2$ is to compute the L2 distance. Via Equation.~\ref{eq:meter}, the average localization error of ResNet1D-[1,1,1,1] on our dataset is 0.0904$m$, which is accurate enough considering typical smart device distribution in homes.

Besides, we would like to report the average mis-classified error between the prediction the the ground-truth, once one sample is incorrectly predicted, via 
\begin{equation}\label{eq:meter}
AME = \frac{1}{M}\sum_{i=1}^{M}\left \| Coor(P_i) -Coor(GT_i))  \right \|_2^2
\end{equation}
where $M$ is the number of samples that wrongly classified~($M=12$); other symbols keep the same meanings as Equation.~\ref{eq:meter}. Then $AWE$ is 2.0943$m$, between two and three experimental locations in FIGURE.~\ref{fig:location}.

\subsection{Data Visualization}

\begin{figure*}[t]
    \centering
    \includegraphics[width=1\linewidth]{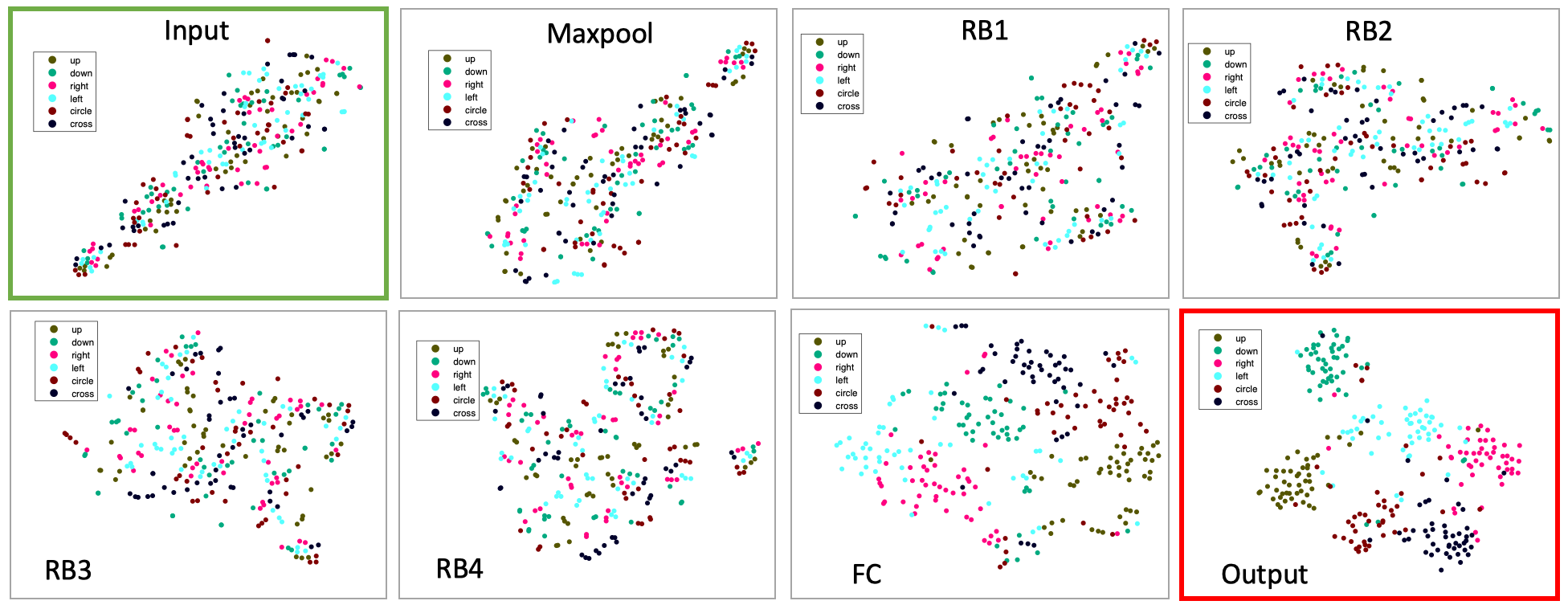}
    \caption{Inputs and feature maps visualization of activity recognition by t-SNE.}
    \label{fig:tsne_act}
\end{figure*}

\begin{figure*}[t]
    \centering
    \includegraphics[width=1\linewidth]{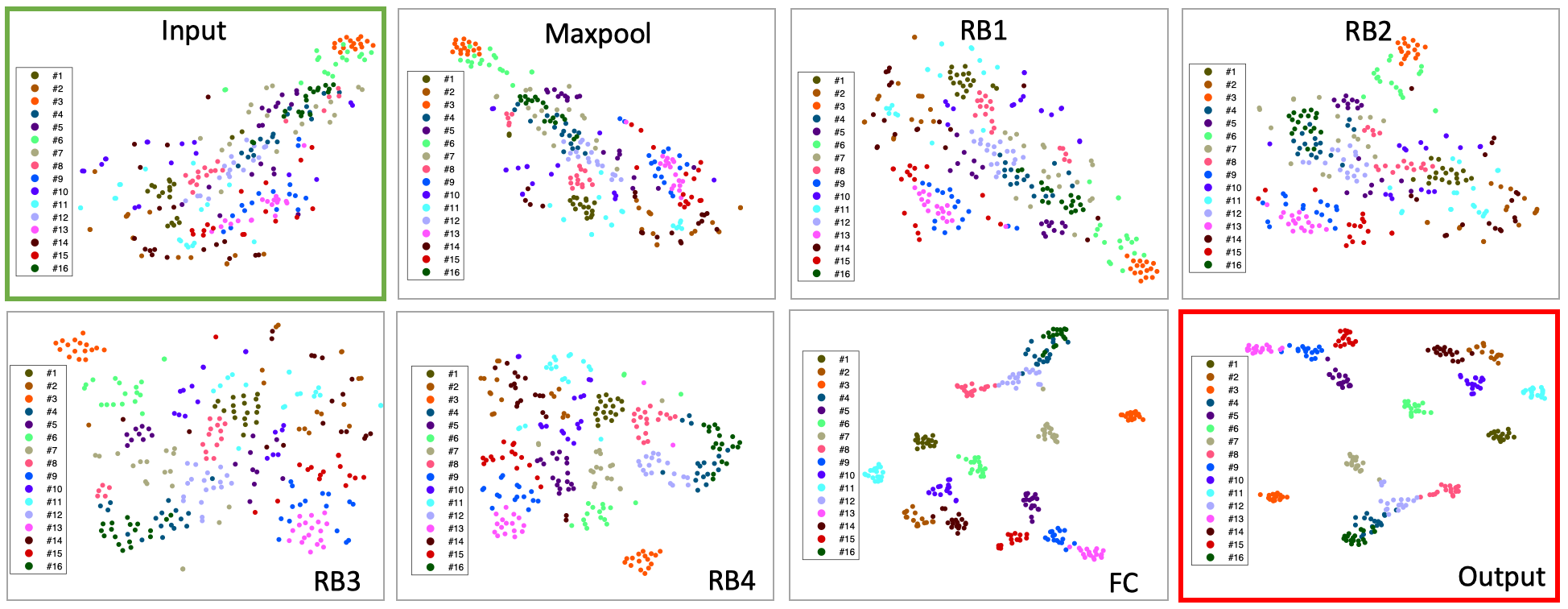}
    \caption{Inputs and feature maps visualization of  indoor localization by t-SNE.}
    \label{fig:tsne_loc}
\end{figure*}

We visualize the test set by t-SNE~\cite{maaten2008visualizing} to explore the behaviors of ResNet1D-[1,1,1,1] on the joint task. Taking the activity recognition task as an example (FIGURE~\ref{fig:tsne_act}), we reduce the input into 2 dimensions data by a t-SNE tool\footnote{\url{https://lvdmaaten.github.io/tsne/}} and display the 2-d data in the figure.
For an input test sample, the reducing procedure is as follows. As~\ref{sec:1d} said, one original CSI fingerprint $C \in R^{52 \times t}$, where $t$ is 192  after the cutting and linear interpolating preprocess, described in~\ref{sec:preprocess}. However t-SNE requires the input to be a long 1D vector, thus we reshape $C$ to be a vector, making a $C'\in R^{1\times 9984}$~($192\times52=9984$). In addition, we repeat the reshaping over all test samples and finally visualize the reshaped samples on the 1st sub-figure in FIGURE~\ref{fig:tsne_act}, marked with the green box. We can see that the inputs are highly disordered in term of activity recognition. 

Besides the raw inputs visualization, we also visualize feature maps produced by ResNet1D-[1,1,1,1] in multiple layers (FIGURE~\ref{fig:network}), i.e., feature maps after max pooling layer, RB1, RB2, RB3, RB4, feature maps before FC~(the 7th sub-figure), and feature maps after FC~(outputs, the 8th-subfigure). We reshape all feature maps to 1D long vectors with the same approach used for visualizing the raw inputs. FIGURE~\ref{fig:tsne_act} shows that ResNet1D-[1,1,1,1] gradually increases the discriminative power of feature maps for the activity recognition task step by step, making classification more accurate in the deeper layers of the network. Finally in the outputs~(the last sub-figure of FIGURE~\ref{fig:tsne_act}), features are learned to be effective for activity recognition. 

With the similar approach, we visualize the raw inputs and feature maps after multiple layers of ResNet1D-[1,1,1,1] for indoor localization in FIGURE~\ref{fig:tsne_loc}. It demonstrates that the network can effectively learn features for indoor location. In FIGURE~\ref{fig:tsne_act}, ResNet1D-[1,1,1,1] generates discriminative features after FC, whereas in FIGURE~\ref{fig:tsne_loc} it generates discriminative features beginning at RB4. Moreover ResNet1D-[1,1,1,1] is able to generate better features for the indoor localization than the activity recognition because the class clusters are more tighter compared the last sub-figure of FIGURE~\ref{fig:tsne_act} and the last sub-figure of FIGURE~\ref{fig:tsne_loc}.

More importantly in the activity recognition, we find the features are largely enhanced through its own branch because the feature before FC~(7th) is much better than the features after the shared RB4~(6th). Under this consideration, we just add one more `C1D$3\times 1$, 512' between the `C1D$3\times 1$, 512' and the `AvgPool1D$4\times 1$', named ResNet1D-[1,1,1,1]+. We train ResNet1D-[1,1,1,1]+, and find it with better performance on activity recognition, listed in Table~\ref{tab:apl_plus}.

\begin{table}[t]
\centering
\begin{tabular}{c|c|c}
\hline
    Model   &  Activity Recognition   & Indoor Localization    \\ \hline\hline
ResNet1D-[1,1,1,] & 88.13\% & 95.68\% \\ \hline
ResNet1D-[1,1,1,1]+ & 89.57\% & 95.68\% \\ \hline
\end{tabular}
\caption{Inspired by t-SNE visualization, we propose  ResNet1D-[1,1,1,1]+, which outperforms ResNet1D-[1,1,1,1] on activity recognition.}
\label{tab:apl_plus}
\end{table}

\begin{table}[t]
\centering
\begin{tabular}{c|c|c|c}
\hline
Method                     & AR    & IL   & Time Cost\\ \hline\hline
DTW+kNN                & 83.45\% & 95.32\% & 3356s\\ \hline
SVM-RBF\cite{CC01a}                    & 40.64\% & 70.32\%& 69s\\ \hline
ResNet1D-[1,1,1,1] & 88.13\% & 95.68\% &161s \\ \hline
ResNet1D-[2,2,2,2] & 87.77\% & 96.40\% & 240s\\ \hline
ResNet1D-[3,4,6,3] & 85.17\% & 97.12\% & 412s\\ \hline
\end{tabular}
\caption{Expansible study and comparison with DTW+KNN. Deeper networks work better on indoor localization task, wheres work worse on activity task. All ResNet1Ds outperform the baseline methods, i.e., DTW+kNN and SVM~\cite{CC01a}. `AR' and `IL' are abbreviations of activity recognition and indoor localization, respectively.}
\label{tab:comparison}
\end{table}

\subsection{Expansible study and Baselines}

ResNet1D is expansible by simply customizing the number of residual block in each RBs, shown in FIGURE~\ref{fig:network}. Following the default settings in ResNet~\cite{he2016deep}, we evaluate the accuracy of ResNet1D-[2,2,2,2] and ResNet1D-[3,4,6,3]. As listed in Table~\ref{tab:comparison}, all ResNet1Ds work well in joint activity recognition and indoor localization. Meanwhile it deserves mentioned that deeper ResNet1Ds tend to work better on indoor location, whereas work worse on activity recognition.

Besides in Table~\ref{tab:comparison}, we show the comparison between ResNet1Ds and two baseline methods, Dynamic Time Warping\footnote{\url{https://www.mathworks.com/matlabcentral/fileexchange/43156-dynamic-time-warping-dtw?focused=3846333&tab=function}}~(DTW)+kNN~\cite{berndt1994using,bagnall2017great}, and Support Vector Machine~(SVM)~\cite{CC01a} with radial basis function  kernel~(RBF). All our proposed ResNet1Ds outperform the baseline DTW+kNN, and SVM-RBF. We also record the time cost of these methods, which are the sum of time cost in training and testing. SVM-RBF costs least but performs worst. DTW+kNN is a very strong baseline in time-serial classification, however it is time-consuming\footnote{Code is available at~\url{https://github.com/geekfeiw/apl}}.

\section{Conclusion}
In this paper, we propose novel 1D Convolutional Neural Networks~(C1D) with two branches for the joint task of activity recognition and indoor localization with WiFi fingerprints. To evaluate the proposed network, we implement IEEE 802.11n protocol in a software-defined-radio hardware, Etuss N210, collect a dataset mainly for human-interaction applications, and fully discuss the results in various aspects.
Experiment results show that our proposed network can achieve joint activity recognition and indoor localization well, e.g., 88.13\% on activity recognition and 95.68\% on indoor localization,   outperforming Dynamic Time Warping and Support Vector Machine. The computational time cost is larger than Support Vector Machine, but can be reduced by training with the larger batch size.
Further, we envision C1D to be one of leading approaches in CSI fingerprint processing, to this end, we release our data and code.

\bibliographystyle{IEEEtran}
\bibliography{reference}

\clearpage

\EOD

\end{document}